\documentclass[aps,twocolumn,floatfix,superscriptaddress]{revtex4-1}
\usepackage{bm}
\usepackage{amssymb}
\usepackage{amsfonts}
\usepackage{graphicx}
\usepackage{color}
\usepackage{longtable}

\begin{document}

\title{First-principles characterization of ferromagnetism in N-doped SrTiO$_3$ and BaTiO$_3$}

\author{Kesong Yang}
\affiliation{School of Physics, State Key Laboratory of Crystal Materials, Shandong University, Jinan 250100, China}
\email{kesong.yang@gmail.com}
\affiliation{Department of Mechanical Engineering and Materials Science, Duke University, Durham, North Carolina 27708, USA}
\author{Ying Dai}
\affiliation{School of Physics, State Key Laboratory of Crystal Materials, Shandong University, Jinan 250100, China}
\author{Baibiao Huang}
\affiliation{School of Physics, State Key Laboratory of Crystal Materials, Shandong University, Jinan 250100, China}

\begin{abstract}
The spin-polarization and magnetic coupling character of N-doped SrTiO$_3$ and BaTiO$_3$ are studied through first-principles calculations. The substitutional N doping at O sites leads to a half-metallic property and produces a magnetic moment of 1.0 $\mu_B$. The magnetic interaction between the nearest and next-nearest N dopants results in a strong ferromagnetic coupling. When the distance between the N dopants is larger than 7 \AA, the ground state of the system tends to be paramagnetic. A nitrogen-concentration threshold to produce the ferromagnetism is estimated. The calculated results give a good explanation for the experimentally observed ferromagnetism in N-doped SrTiO$_3$ and BaTiO$_3$.
\end{abstract}
\maketitle

Diluted magnetic oxides have attracted much attention owning to their fundamental physics and promising applications in the spintronics.\cite{Dietl_Nat.Mater_2010} With respect to the
transition-metal doped oxides, an unexpected room-temperature ferromagnetism (RTFM) has recently been observed in C-doped ZnO.\cite{Pan_PRL_2007} Soon after, lots of first-principles theoretical
calculations have been done to predict the possible ferromagnetism in C (N)-doped oxides, such as ZnO,\cite{Shen_PRB_2008,Yang_PRB_2010} TiO$_2$\cite{Yang_APL_2008, Yang_CPL_2009, Tao_APL_2009} and SnO$_2$,\cite{Rahman_APL_2010,Xiao_SSC_2009} and the RTFM in these C (N)-doped
oxides was confirmed by the later experiments.\cite{Wen_IEEE.Trans.Magn_2009,Ye_PLA_2009,Cruz_JPCM_2009,Bao_JAP_2011,Hong_APL_2011}
Recently, the RTFM was found in N-doped SrTiO$_3$ (STO)\cite{Liu_CJP_2009} and BaTiO$_3$ (BTO).\cite{Tan_JAC_2011} It is well-known that the perovskite STO and BTO are
important ferroelectric materials,\cite{Dawber_PRL_2005} and thus it is of great interest to realize the coexistence
of ferroelectricity and ferromagnetism in the two oxides for their multiferroic applications.\cite{Ramesh_Nat.Mater_2007}
Liu et al.'s experimental studies on the N-doped STO show that the ferromagnetism only
appears in the sample with a relatively high doping concentration, and N doping leads to an
insulator-metal conducting phase transition.\cite{Liu_CJP_2009} As a result, it is speculated that there might
 exist a nitrogen-concentration threshold, only above which can the N-doped STO and BTO
 exhibit the long-range ferromagnetism. To address this question and understand the origin of its
 ferromagnetism, we studied the spin-polarization and magnetic coupling character of N-doped STO
 and BTO on the basis of the first-principles calculations. The theoretical results show that the
 N doping leads to the typical half-metallic properties, and produces a strong ferromagnetic coupling
through the magnetic orbital interaction between the nearest and next-nearest N dopants.
A nitrogen-concentration threshold to produce the long-range ferromagnetism is also estimated.

Various doped structures with N atoms at O sites are modeled using the $3\times 3\times 3$ supercell
containing 135 atoms based on the cubic perovskite STO and BTO, as shown Fig. 1. The projector augmented wave (PAW) potentials are used for electron-ion interactions and generalized gradient approximation parameterized by Perdew-Burke-Ernzerhof (PBE) is used for exchange-correlation functional.\cite{Perdew_PRL_1996} The cut-off energy of 400 eV for the plane-wave basis set and a $2\times 2\times 2$ k-point grid centered at $\Gamma$ point are used.\cite{Monkhorst_PRB_1976} The convergence threshold for self-consistent iteration is set at 10$^{-6}$ eV, and all the atomic positions are fully optimized until all components of the residual forces are smaller than 0.01 eV/\AA. All the calculations are performed using the Vienna ab-inito simulation package (VASP).\cite{Kresse_PRB_1996,Kresse_CMS_1996}

\begin{figure}
\center
\includegraphics[scale=0.2]{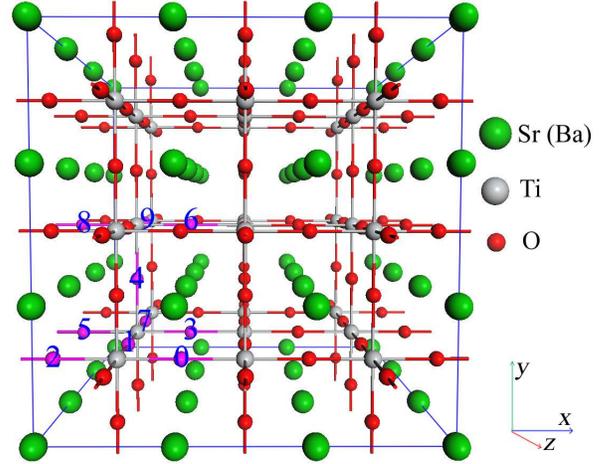}
\caption{(Color online) $3\times3\times3$ supercell employed to simulate
N-doped STO and BTO. The O atoms in pink color labeled by 0-9 are the sites to be replaced with N atoms.
The {\emph{x-}}, {\emph{y-}} and {\emph{z-}} axes are along the crystallographic {\emph{a-}}, {\emph{b-}} and {\emph{c-}} directions, respectively.}
\end{figure}

To understand the origin of the spin-polarization in N-doped STO and BTO, we firstly studied one-N-atom doped model, which is constructed by replacing one O atom at the 0 site using one N atom in the supercell (see Fig. 1). The calculated total density of states (TDOS) and partial DOS (PDOS) for N-doped BTO and STO are shown in Fig. 2a (a$^\prime$) and 2b (b$^\prime$), respectively. It shows that the substitution of N for O introduces some spin-polarized impurity states in the band gap, and they mainly originate from N 2p orbitals. The Fermi level is pinned in the middle of these band-gap states, indicating that N-doped STO and BTO exhibit the typical half-metallic character. This is good agreement with the previous theoretical calculation\cite{Bannikov_JMMM_2008} and Liu et al.'s experiment that N-doped STO shows the conducting property.\cite{Liu_CJP_2009} With respect to the local Cartesian coordinate defined in Fig. 1, the up-spin and down-spin N 2{\emph{p$_x$}} orbitals are occupied, and the up-spin N 2{\emph{p$_y$}} and 2{\emph{p$_z$}} orbitals are also occupied. However, the Fermi level is pinned in the middle of the down-spin N 2\emph{p$_y$} and 2\emph{p$_z$} orbitals, i.e., half of the down-spin N 2\emph{p$_y$} and 2\emph{p$_z$} orbitals are occupied. This indicates that one N atom at an O site generates a magnetic moment of 1.0 $\mu_B$, and the N dopant exists as a N$^{2-}$ (\emph{s$^2$p$^5$}) anion. In addition, non-spin-polarized calculations for one-N-atom doped STO and BTO show that the spin-polarized state is more stable than the non-spin-polarized state by about 124 and 99 meV, respectively, thus indicating that the ground state of N-doped STO and BTO is magnetic.

\begin{figure}
\centering
\includegraphics[scale=0.41]{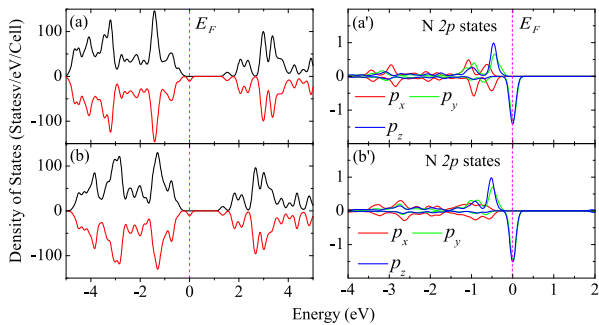}
\caption {(Color online) The total DOS and partial DOS plot for one-N-doped (a) STO and (b) BTO. The dotted line indicates the Fermi level at 0 eV.}
\end{figure}

To examine the type and strength of magnetic coupling between the two N dopants, nine inequivalent structural
configurations are modeled using the 135-atom $3\times 3\times 3$ supercell, which are
constructed by replacing two O atoms using two N atoms at positions (0, 1), (0, 2), (0, 3), (0, 4), (0, 5),
(0, 6), (0, 7), (0, 8) and (0, 9), respectively, as shown in Fig. 1. For convenience, we use (i, j)
to denote the structure in which the O atoms at the positions (i, j) are replaced by N atoms. To examine the
stable magnetic coupling type, both the ferromagnetic (FM) and the antiferromagnetic (AFM) coupling between the
moments of two N dopants are considered for each structure. In TABLE I and II, we summarize the calculated results
for N-doped STO and BTO. For each structure, the optimized N$\cdots$N distance and the relative energy $\Delta E$,
the magnetic energy \emph{E$_{mag}$}=\emph{E$_{FM}$}-\emph{E$_{AFM}$}, the total magnetic moment (\emph{M}) under the ferromagnetic alignment and the
calculated Curie temperature (\emph{T$_C$}) are listed. For the (i, j) configuration, the relative energy $\Delta E$ is defined as
the energy difference between the energy of the lower-energy (either FM or AFM) and the total energy of the (0, 1)
structure, i.e., $\Delta E$=\emph{E}(i, j)-\emph{E}(0, 1). The \emph{T$_C$} is calculated based on the mean-field theory and Heisenberg model, i.e., \emph{k$_B$T$_C$}=(2/3)\emph{E$_{mag}$}.\cite{Kudrnovsky_PRB_2004,Maca_APL_2008} Here \emph{E$_{mag}$} is the magnetic energy.

\begin{table}
\centering
\caption{Calculated results for N-doped STO: the optimized N$\cdots$N distance (d$_{N-N}$) (\AA), the relative energies $\Delta E$ (eV), magnetic energy \emph{E$_{mag}$}=\emph{E$_{FM}$}-\emph{E$_{AFM}$)} (meV), the total magnetic moment ($M$) ($\mu_B$/cell) under ferromagnetic alignment and Curie temperature (\emph{T$_C$}) (K) calculated for the $(i,j)$ structure of the two-N-atom doped STO.}
\begin{ruledtabular}
\begin{tabular}{cccccc}
$(i,j)$  & $d_{N-N}$ & $\Delta E$ & $E_{mag}$ & $M$ & $T_C$  \\
\hline
(0, 1)	&	2.66	&	0	&	-144	&	2	&	1108	 \\
(0, 2)	&	3.99	&	0.44	&	-114	&	2	&	 877	 \\
(0, 3)	&	3.93	&	0.26	&	-141	&	2	&	 1085	 \\
(0, 4)	&	4.8	&	0.2	&	-2	&	2	&	15	 \\
(0, 5)	&	5.51	&	0.27	&	-109	&	2	&	 838	 \\
(0, 6)	&	5.52	&	0.35	&	-7	&	2	&	54	 \\
(0, 7)	&	6.22	&	0.22	&	-1	&	2	&	8	 \\
(0, 8)	&	6.82	&	0.42	&	-7	&	2	&	54	 \\
(0, 9)	&	7.37	&	0.19	&	0	&	2	&	0	 \\
\end{tabular}
\end{ruledtabular}
\end{table}

\begin{table}
\centering
\caption{Calculated results for N-doped BTO.}
\begin{ruledtabular}
\begin{tabular}{cccccc}
$(i,j)$  & $d_{N-N}$ & $\Delta E$ & $E_{mag}$ & $M$ & $T_C$  \\
\hline
(0, 1)	&	2.77	&	0	&	-213	&	2	&	1638	 \\
(0, 2)	&	4.11	&	0.47	&	-50	&	2	&	385	 \\
(0, 3)	&	4.02	&	0.24	&	-111	&	2	&	 854	\\
(0, 4)	&	4.95	&	0.3	&	-2	&	2	&	15	\\
(0, 5)	&	5.67	&	0.29	&	-28	&	2	&	215	 \\
(0, 6)	&	5.66	&	0.3	&	-6	&	2	&	46	\\
(0, 7)	&	6.38	&	0.5	&	2	&	2	&	$-$	\\
(0, 8)	&	6.99	&	0.17	&	0	&	2	&	0	 \\
(0, 9)	&	7.56	&	0.18	&	0	&	2	&	0	 \\
\end{tabular}
\end{ruledtabular}
\end{table}

  Table I and II both show that the relative energies \emph{$\Delta$E} of various (i, j) configurations are nearly uniform, indicating that the N dopants in STO and BTO do not have the energetic preference to form a cluster. For the (0, 1), (0, 2) and (0, 3) structures, a strong FM coupling occurs between the magnetic moments of two N dopants. This is consistent with the experimentally observed ferromagnetism in N-doped STO\cite{Liu_CJP_2009} and BTO.\cite{Tan_JAC_2011} Furthermore, the highest calculated \emph{T$_C$} of N-doped STO and BTO both exceed 1000K, much higher than that of C-doped ZnO (about 400 K).\cite{Pan_PRL_2007, Yang_PRB_2010} Therefore, N-doping is an effective route to obtain high-\emph{T$_C$} half-metallic ferromagnetism in STO and BTO. For (0, 5) structure, a stronger FM coupling takes place in N-doped STO than in doped BTO. As the N$\cdots$N distance increases, i.e., for the (0, 4), (0, 6), (0, 7), (0, 8) and (0, 9) structures, their magnetic energies decrease a lot, and even become zero when the N$\cdots$N distance exceeds about 7\AA. This indicates that N-doped STO and BTO tend to be paramagnetic when the distance between the two N dopants is long enough. It means that an effective magnetic coupling interaction only occurs between the two nearest and next-nearest neighbor N dopants. As a result, we speculate that there exists a percolation threshold of nitrogen-concentration,\cite{Osorio_PRL_2006} and if the magnetic interaction range is limited to the next-nearest N dopants, a minimal nitrogen-concentration of about 11.1 \% is essential to produce the long-range ferromagnetism. This gives a good explanation for the experimental fact that the ferromagnetic hysteresis loop only appears in N-doped STO with a relatively high concentration (see Fig. 4 in Ref.\cite{Liu_CJP_2009}).

\begin{figure}
\centering
\includegraphics[scale=0.2]{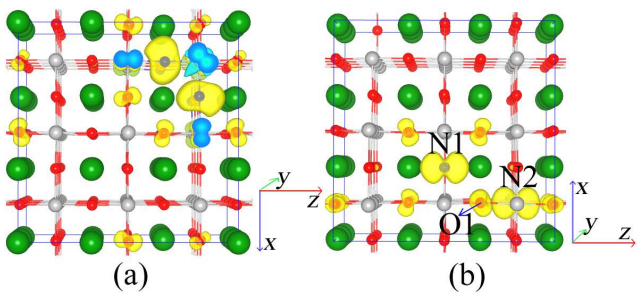}
\caption {(Color online) Calculated spin density distribution of two-N-atom doped STO for (a) (0, 1) configuration and (b) (0, 4) configuration.}
\end{figure}

To understand the ferromagnetic coupling mechanism in N-doped STO and BTO, the (0, 1) and (0, 4) structures of N-doped STO are chosen as examples to show their three-dimensional spin density distributions, which are presented in Fig. 3. The spin density is mostly contributed by the N dopants as well as their nearest O atoms. For (0, 1) structure, the magnetic orbitals of the two N dopants have a substantial direct overlap, while for (0, 2), (0, 3) and (0, 5) structures, though not shown here, an indirect magnetic orbitals overlap through their common O atoms is found. Therefore, a direct or indirect magnetic orbitals overlap between the two N sites could be responsible for the FM coupling.\cite{Yang_APL_2008} With respect to the other remaining configurations, the N$\cdots$N distance in (0, 4) structure is much shorter, and to clearly indicate the magnetic coupling interaction between two N dopants, its spin density distribution is shown in Fig. 3b. In this configuration, the two N dopants induce the spin-polarization of the nearest O atoms in the different plane (N1 is in \emph{xz} plane, and N2 in \emph{yz} plane). Hence, the magnetic orbitals of the two N dopants could only have a weak overlap through only one common O atom (labeled as O1 in Fig. 3b), thus resulting in a weak magnetic interaction. For the other structures, the long N$\cdots$N distances determine that the magnetic orbital overlap between the two N dopants is small or even zero, and thus their magnetic ground states tend to be paramagnetic.

In summary, our first-principles calculations for N-doped STO and BTO indicate that the N doping introduces spin-polarized impurity states in the band gap, generating a magnetic moment of 1.0 $\mu_B$, and leads to the half-metallic property. The substantial magnetic interaction between the nearest and next-nearest N dopants leads to the strong ferromagnetic coupling, but the ground state tends to be paramagnetic when the distance between the N dopants is larger than about 7 \AA. A minimal nitrogen-concentration to produce long-range ferromagnetism is estimated. These results give a good explanation for the experimental phenomena, and provide a possible route to realize their multiferroic applications.
\begin{acknowledgments}
This work is supported by the National Basic Research Program of China (973 program, 2007CB613302), National Science foundation of China under Grant 11174180 and 20973102, and the Natural Science Foundation of Shandong Province under Grant number ZR2011AM009.
\end{acknowledgments}


\end{document}